# Mechanical Controlled Thermal Switch and Hysteresis with Domain Boundary Engineered Phonon Transport

*Suzhi Li[1], Xiangdong Ding[1,2*], Jie Ren[2*], Ju Li[1,3], Jun Sun[1] and Ekhard K.H. Salje[1,4*]*

1. State Key Laboratory for Mechanical Behavior of Materials, Xi'an Jiaotong University, Xi'an 710049, China
2. Theoretical Division, Los Alamos National Laboratory, Los Alamos, New Mexico 87545, USA
3. Department of Nuclear Science and Engineering and Department of Materials Science and Engineering, Massachusetts Institute of Technology, Cambridge, Massachusetts 02139, USA
4. Department of Earth Sciences, University of Cambridge, Cambridge CB2 3EQ, UK

*E-mail:   dingxd@mail.xjtu.edu.cn

renjie@lanl.gov

ekhard@esc.cam.ac.uk

**Abstract**

Heat flow control in phononics has received significant attention recently due to its widespread applications in energy transfer, conversion and utilization. Here, we demonstrate that by applying external stress or strain we can effectively tune the thermal conductivity through changing the density of twin boundaries, which in turn offers the intriguing mechanical-controlled thermal switch and hysteresis operations. Twin boundaries perpendicular to the transport direction strongly scatter phonons. As such, the heat flow is in inverse proportional to the density of twin boundaries and hence allows an excellent way to switch thermal conductivity mechanically and even leads to the interesting hysteresis behavior as a thermal memory. Our study provides a novel mechanism to couple thermal and mechanical properties of materials as a matter of "domain boundary engineering" and can have substantial implications in flexible thermal control and thermal energy harvesting.

Heat conduction and electric conduction are two fundamental energy transport phenomena in nature. The development of modern electronics to control charge transfer has impacted every aspect of our everyday life. However, heat conduction has never been treated on the equal footing and its flexible control is largely absent, in spite of the fact that almost 90 percent of the world's energy are utilized through heat processes.[1] To meet the demand, a new discipline, phononics emerges,[2] which is the science and technology of phonons that are main carriers of energies in non-metals. Phononics aim to manipulate phonons and to control heat flow and thermal energy as flexibly as in electronics. To achieve this ultimate goal, various functional thermal devices, like thermal diodes, thermal transistors, thermal logic gates have been proposed theoretically and partially been realized in experiments.[2] It was demonstrated that phononic logic gates can perform similar operations as their electronic counterparts which makes it possible that phonons can be even used as the information carriers.[3] In this sense, thermal memory materials[4] can be designed if the micro-structural conditions for functional materials are better understood. Nevertheless, candidate materials for potential applications have yet to be identified. Moreover, the controls of those phononic devices are mainly based on tuning temperatures. More flexible manipulating protocols are in demand. Considering that microstructure changes can tune the lattice vibrations, we can effectively manipulate the microstructures to control the phonon fluxes so as to generate diverse device applications based on the heterogeneous phonon properties of a material.

It is the purpose of this paper to show that by mechanically governing the microstructures rather than the intrinsic properties of the single domain state, it is possible to generate two logic states, one with high heat conduction and the other with low heat conduction, for thermal information storage and thermal switch. This idea is intuitive since it is known that phonon fluxes are strongly influenced by internal defects. Among all the types of defects, such as point defects, line defects, etc, interfacial boundaries are the ones that can best influence the thermal transport and are susceptible to external fields such as elastic stresses and strains in ferroelastics,[5] electric fields in ferroelectrics[6] and magnetic fields in ferromagnets. Mobile interfaces will, in each case, modify the phonon scattering and hence allows the modulation of the heat flux. For instance, recent experiment shows that the conductivity of Si-Ge alloy thin films is 3 ~ 5 times lower than the bulk values.[7] The reason lies in that long wavelength phonons are scattered strongly in film boundaries. Previous research also focused on the thermal properties of other types of interfaces,[8-10] such as bi-crystal interfaces, tilt grain boundaries, and free surfaces. None of the above mentioned interfaces can be tuned as we propose in this paper.

We investigated coherent twin boundaries because they are most easily manipulated mechanically. Our recent work has already shown that the morphologies and properties of twin patterns (TPs) could be changed greatly by mechanically deformation,[11-13] which leads us directly towards ways to tune thermal conductivity via optimizing twin patterns and further develop thermal memory devices. The important roles of domain boundaries on materials properties have already shown in many other fields,[14-16] which is known as "domain boundary engineering".[17] For example, weak doping generates superconducting twin walls below 3K in $WO_3$ while the matrix is non-superconducting.[14] In ferroelastic systems, certain alloys can take novel shape memory behaviors with the aid of domain boundary motion.[18] The interaction of domain wall and point defects largely determines the lifetime of ferroelectric memory devices,[19] while ferroelectric domain boundaries were found in paraelectric bulk materials.[20]

We will show in this paper that within the same framework of "domain boundary engineering",[17] it is possible to manipulate the thermal conductivity and obtain two stable states for information storage by optimizing domain structures and domain patterns. We studied the effect of mechanically driven domain boundaries evolution on heat transfer by using non-equilibrium molecular dynamics (NEMD) technique.[21] We show that thermal conductivity is strongly related to the morphology of twin patterns. By applying different strain/stress state, we can effectively manipulate morphology of twin structure and eventually tune thermal conductivity by a factor 2 ~ 4.

We used a two-body potential to represent the interactions of atoms in two-dimensional (2D) system. The potential energy $U(r)$ contains three parts, as the first-nearest atomic interactions of $20(r - 1)^2$, the second-nearest interactions $-10(r - \sqrt{2})^2 + 2000(r - \sqrt{2})^4$ and third-nearest interactions $-(r - 2)^4$, where $r$ is atomic distance. This potential is developed based on Landau theory by choosing the shear angle as "order parameter", and this model is generic to all ferroelastic materials for studying twinning and mobility of twin boundaries. The details of properties obtained by this potential are described in our previous work.[11-13, 15] The equilibrium unit cell is in shape of parallelogram with the shear angle of 4 degrees. To correspond a relatively real system, we set the equilibrium lattice constant $a = 1$ Å and atomic mass $M = 100$ amu. We construct a sandwich twinned structure containing two horizontal twin boundaries (HTBs), as shown in **Figure 1**. The surface ratio of intermediate layer to the whole sample is fixed to be 0.7. The size of 2D simulation box is $160a \times 100a$ (= 16 nm × 10 nm) in $xy$ plane, where $a$ (= 1 Å) is the lattice repetition unit. Periodic boundary condition is applied along the $x$ direction and free boundary condition is used in the $y$ direction. To study the structure evolution upon the external load, the top and bottom several layered atoms were fixed rigidly as the loading grip. We applied different kinds of strain state to examine the twinning evolution in mechanically driven system. The strain tensor in 2D system can be described as $[\varepsilon_{xx}, \varepsilon_{yy}, \gamma_{xy}]^T$. In our work, the deformation was performed at different $\varepsilon_{yy}/\gamma_{xy}$ ratio with $\varepsilon_{xx} = 0$. The dynamic loading was taken at $T = 70$ K ($\approx 0.2\ T_m$, where $T_m$ is melting point) by using Nosé-Hoover thermostat.[22, 23] Under such low temperature, the diffusion process can be greatly suppressed. All the calculations are performed using the LAMMPS code.[24]

For each loading step, we used the NEMD technique to calculate thermal conductivity in a given configuration.[21] The idea is to apply a heat flux in the system along the $x$ direction, which will result in a temperature gradient. When the heat transfer becomes a steady flow, the thermal conductivity $\kappa$ along $x$ direction is calculated via Fourier law as $\kappa = -J_x/(\partial T/\partial x)$, where $J_x$ is the heat flux from heat source to heat sink and $T$ is temperature. After applying heat transfer, the induced temperature gradient will range from 101 K (= 0.28 $T_m$) to 47 K (= 0.13 $T_m$), corresponding to the heat source and the heat sink (see Fig. S1 in Supplementary Information).

We first apply a simple shear deformation to the sample with strain tensor as $[0, 0, \gamma_{xy}]^T$. **Figure 2**a shows the response of shear stress with shear strain ($\gamma_{xy}$) in a cycle. For the loading process (see black curve), the sample yields when $\gamma_{xy}$ reached 0.6%. The sample was unloaded to zero-strained state from $\gamma_{xy} = 1.6\%$ (see red curve). Figure 2b shows the variation of $\kappa$ during the loading/unloading loop. In elastic regime, the magnitude of $\kappa$ does not change significantly ( ~ 140 to 150 W/m/K). When loading into the plastic regime, $\kappa$ undergoes an abrupt drop to ~ 90 W/m/K, almost one half of the initial value. This magnitude is sustained under further stress. When the system is unloaded, $\kappa$ increases again and shows interesting hysteresis behavior during the loading/unloading cycle. This hysteresis of thermal transport could lead to the novel thermal memory device application[4, 25] and is based on the underlying twin boundary dynamics.

We examined the evolution of twin pattern for several typical strains (marked as (c), (d), (e), (f) in Figure 2b) upon loading/unloading, as shown in Figure 2c-2f. When the system deforms plastically, new horizontal twin layers are induced by the applied shear strain, accompanied with the formation of a certain amount of vertical twin boundaries (VTBs). VTBs nucleate from one horizontal twin boundaries (HTBs), propagate and terminate in another horizontal twin boundary. The new-formed VTBs and HTBs superimpose and finally evolve into a much more complicated twin pattern. This pattern then shows much reduced thermal conductivity. Moreover, the structure of twin pattern is quite stable and the twinning morphology will not undergo large changes even under the applied temperature gradient in NEMD calculations (see Fig. S2 in Supplementary Information). Upon unloading, horizontal twins will be preserved and produce permanent deformation of the sample. Vertical twins are unstable and will vanish gradually when removing the external load, which leads to an increase of $\kappa$.

These results indicate that it is very likely that the reduction of thermal conductivity results from the existence of vertical twins. To further probe the exact role that VTBs and HTBs play in thermal transfer, we examined the variation of both VTBs density ($\rho_{VTB}$) and HTBs density ($\rho_{HTB}$) in loading/unloading cycle. The units of $\rho_{VTB}$ and $\rho_{HTB}$ are $a^{-1}$. Figure 2g shows clearly that the production and annihilation of vertical twins are highly correlated with changes of the thermal conductivity, i.e., the formation of VTBs reduces thermal transport. The presence of horizontal twins does not seem to influence the heat transport and thermal conductivity is independent of the density of HTBs (Figure 2h). In addition, the dominant effects of VTBs, but not the HTBs, on the reduction of thermal conductivity can be seen in the phonon density of states. We found that the existence of VTBs can severely suppress the contribution of some medium-frequency phonons on the heat transport due to the twin boundary-phonon scattering (see Fig. S3 in Supplementary Information). To distinguish the two twin structures with different heat transfer properties, we term the systems only containing HTBs as twinning pattern 1 (TP1) and that containing VTBs (purely vertical twin boundaries or mixed twin structure) as twinning pattern 2 (TP2). For the present case, the magnitude of $\kappa$ in TP2 can be lowered twice compared with TP1.

The VTBs density is now taken as "field" for tuning thermal properties. We find that the applied strain/stress state effectively determines the mechanically induced density of VTBs in TP2.

When replacing the simple shear to a mixture of loading state [0, $\varepsilon_{yy}$, $\gamma_{xy}$]$^T$, i.e., applying certain amount of tensile strain ($\varepsilon_{yy}$) along y direction besides shear strain ($\gamma_{xy}$), the density of VTBs can be enhanced greatly. **Figure 3**a and 3b show the variation of thermal conductivity and VTBs density with applied shear strain under different $\varepsilon_{yy}/\gamma_{xy}$ ratio, respectively. The VTBs density increases continuously when a tensile strain is applied. Correspondingly, the thermal conductivity is reduced dramatically. Figure 3c-3e show the atomic configurations at $\gamma_{xy}$ = 1.8% under different $\varepsilon_{yy}/\gamma_{xy}$ ratios. We find that the application of tensile strain effectively suppresses the formation of HTBs.

In kinetics theory of phonon transport, the thermal conductivity is proportional to the mean free path of phonons as $\kappa = \Lambda C v/3$, where $\Lambda$ is the mean free path, $C$ is the heat capacity and $v$ is the acoustic velocity. In our system, the mean free path is determined by two kinds of phonon-scattering events, involving (a) phonon-phonon collision and (b) interactions of phonons and domain boundaries. The effective $\Lambda$ can be given by

$$\frac{1}{\Lambda} = \frac{1}{\Lambda_{P-P}} + \frac{1}{\Lambda_{P-TB}} \quad (1)$$

where $\Lambda_{P-P}$ and $\Lambda_{P-TB}$ are lengths of bulk phonon-phonon scattering and phonon-TBs scattering, respectively. Usually, the magnitude of $\Lambda_{P-P}$ approaches some $\mu$m[26, 27] that is much larger than $\Lambda_{P-TB}$ which is restricted to some tenths nm ($\Lambda_{P-P} \gg \Lambda_{P-TB}$). Thus, $\Lambda_{P-TB}$ takes a dominant role in determination of the total mean free path $\Lambda$. For a constant $\Lambda_{P-P}$, $1/\kappa$ should be proportional to $1/\Lambda_{P-TB}$ ($1/\kappa \propto 1/\Lambda_{P-TB}$). On the other hand, $\Lambda_{P-TB}$ here should be equal to the average twin boundary spacing ($\lambda$), which is inverse proportional to the density of VTBs as $\lambda = 1/\rho_{VTB}$. Therefore, we can finally obtain an approximate relationship of $1/\kappa$ and $\rho_{VTB}$ as

$$\frac{1}{\kappa} \propto \rho_{VTB} \quad (2)$$

To quantify this response, we extract the data in the strain rage of 1.2 – 1.8% for different strain state in Figure 3a and plot them in **Figure 4**. The data fit the linear relationship between $1/\kappa$ and $\rho_{VTB}$ in Eq. 2 very well. Therefore, the above equation can be taken as a simple estimate of thermal conductivity that arises from twin boundaries scattering. We have shown that the density of VTBs can be manipulated very precisely by external strains. Since VTBs strongly affect thermal transport, we can hence tune thermal conductivity via the plastic strain of a sample. Although other studies have shown that the existence of interfaces can reduce heat transfer due to phonon scattering,[7] they did not show that this reduction can be used to flexibly tune the thermal conductivity, which is the key requirement for device applications. In this paper, we propose an alternative strategy, namely that thermal conductivity can be tuned by controlling twin patterns and twin boundary densities thought external strain field. The same conclusions will hold for ferroelectric materials where the polar $90^{\circ}$ twin boundaries can be changed by electric fields.

In summary, by using NEMD simulations, we studied the effect of twin patterns evolution on heat transfer in mechanically driven system. We find that the formation of vertical twins can dramatically reduce thermal conductivity. By controlling the applied strain state, we can manipulate the density of vertical twin, which in turn offers us the intriguing mechanical-controlled thermal switch and hysteresis actions. In the framework of "domain boundary engineering", our results give a new avenue to modify thermal conductivity using mechanically-controlled twin boundaries, which will facilitate the development of thermal switches or thermal memory devices for smart energy control and harvesting.

**Acknowledgment**

XD, JS and JL appreciate the support of NSFC (51171140, 51231008, 51321003, 51320105014), the 973 Program of China (2010CB631003, 2012CB619402) and 111 project (B06025). JR acknowledges the support from National Nuclear Security Administration of the U.S. DOE at LANL under Contract No. DE-AC52-06NA25396 through the LDRD Program. EKHS is grateful for support by the Leverhulme fund (RG66640) and EPSRC (EP/K009702/1). Z. Wang is thanked for helpful discussion.

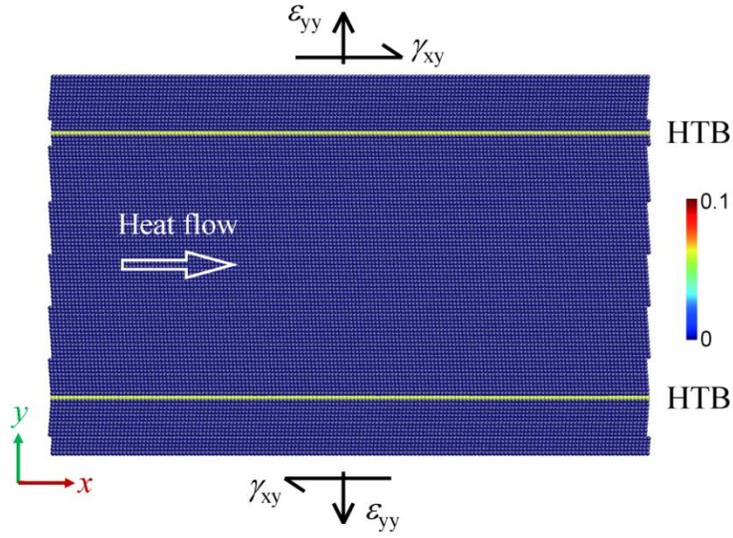

**Figure 1**. A two-dimensional sandwich model with two preexisted horizontal twin boundaries (HTBs). The lattice unit is in shape of parallelogram with tilt angle of 4 degrees. The middle layer has an area ratio of 0.7 to the whole sample. The strain tensor in 2D system can be described as $[\varepsilon_{xx}, \varepsilon_{yy}, \gamma_{xy}]^T$. In our work, the deformation was performed at different $\varepsilon_{yy}/\gamma_{xy}$ ratio with $\varepsilon_{xx} = 0$, where $\varepsilon_{yy}$ is tensile strain along y direction, $\gamma_{xy}$ is shear strain. Heat flow will be applied along $x$ direction. The atom colors are shown according to local symmetry of crystal.

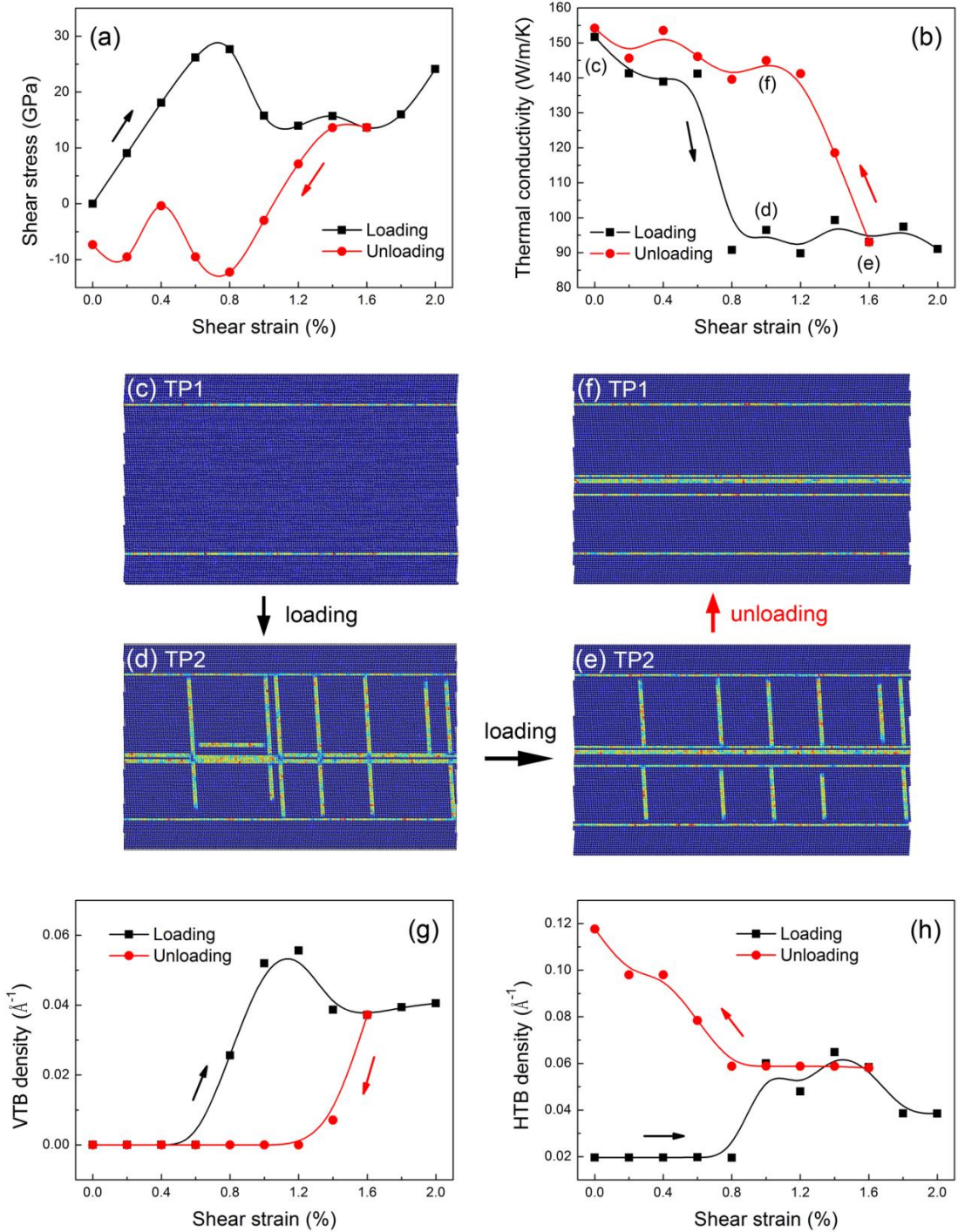

**Figure 2**. Influence of twin pattern evolution on thermal conductivity. (a) Stress-strain curve in a simple shear loop. (b) Variation of thermal conductivity with shear strain. (c)-(e) Typical atomic images marked in (b). Twin pattern 1 (TP1) in absence of vertical twin boundaries (VTBs) has a lower thermal conductivity than twin pattern 2 (TP2) with VTBs. Density of (g) VTBs and (h) HTBs with shear strain in units of $a^{-1}$ (= Å$^{-1}$), where $a$ (= 1 Å) is lattice constant.

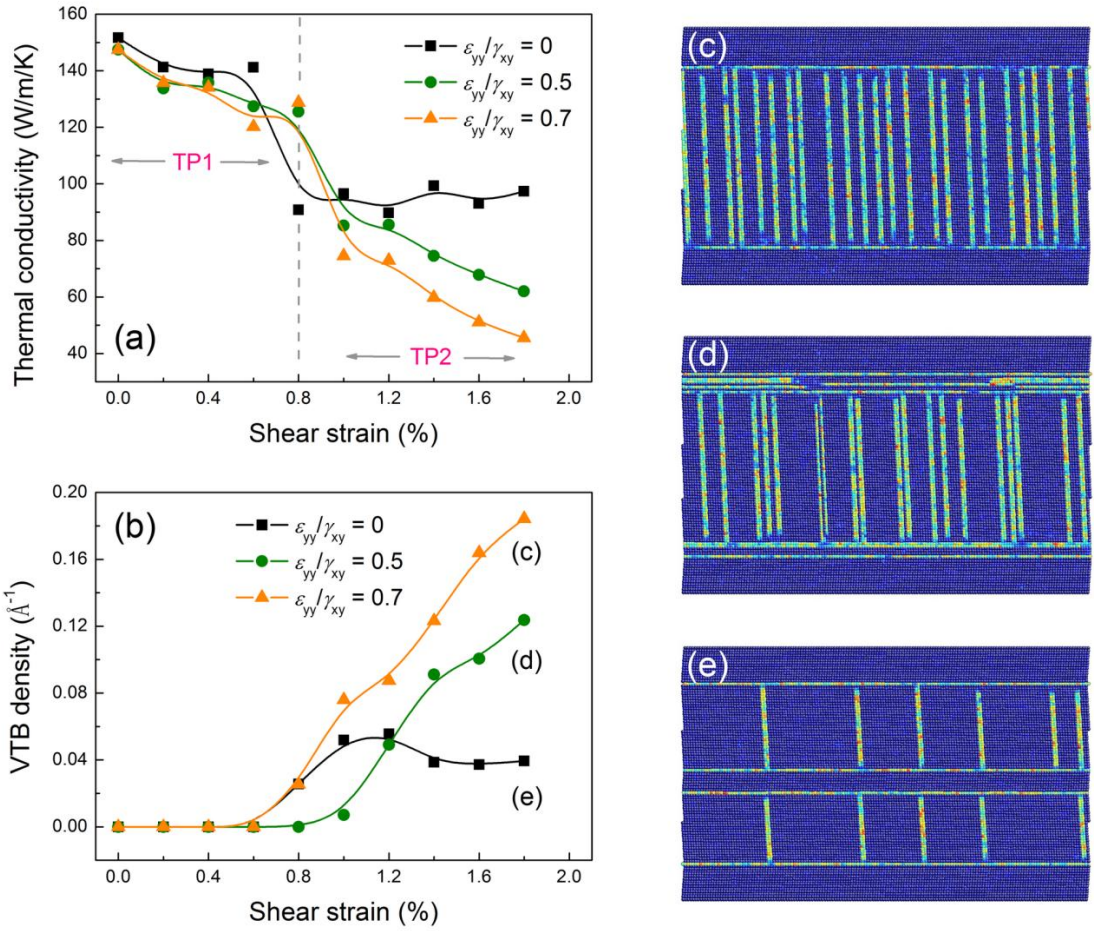

**Figure 3**. Tuning thermal conductivity through manipulating VTBs density by external strain field. (a) Thermal conductivity variation with shear strain under different $\varepsilon_{yy}/\gamma_{xy}$ ratio. (b) Variation of VTBs density with shear strain. (c)-(e) Three typical atomic images marked in (b).

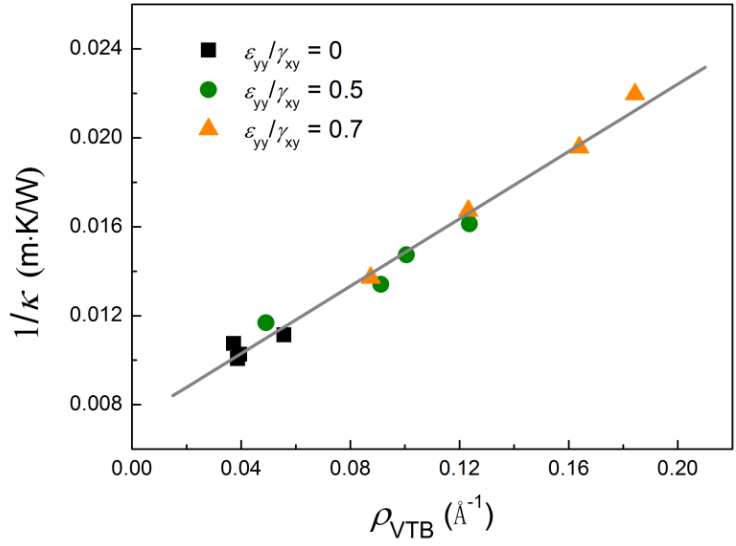

**Figure 4**. The linear response of $1/\kappa$ with VTBs density ($\rho_{\text{VTB}}$). The data correspond to those in Figure 3.

**Supplementary Information**

1. Calculation thermal conductivity by non-equilibrium molecular dynamics (NEMD) simulation

We used NEMD method to calculate thermal conductivity.[1] Specifically, the whole system are divided into 40 slabs along *x* direction (Fig. S1a). The middle and the end slabs are taken as heat sources and heat sinks, separately. The heat flux will be generated by exchanging the kinetic energy between heat source and sinks in a given time frequency until the heat flow reached steady state. Owing to periodic boundary conditions used in x direction, the heat flux will induce two temperature gradients on both sides. Usually, the temperature profile shows non-linear effect in the slabs near the sources and sinks, but behaves linearly in slabs far away from sources and sinks, as shown in Fig. S1b. The thermal conductivity along *x* direction can then be calculated in the linear regime as $\kappa = -J_x/(\partial T/\partial x)$ according to Fourier law, where heat flux $J_x$ ($=\Delta E_K/\tau A$) is the amount of energy exchange ($\Delta E_K$) in a given time ($\tau$) through the cross-sectional area (*A*) and *T* is temperature. Here, since the two-dimensional simulations are performed here, the magnitude of cross-sectional area *A* equal to the box length in *y* direction by setting the length along *z* direction as an unity.

2. The slight effect of temperature gradient on the stability of twin patterns

For calculating thermal conductivity in NEMD method, there will induce a temperature gradient on system. The range is dependent on the frequency of kinetic exchange between heat sources and sinks. In our simulation, the dynamics loading is performed at $T = 70$ K (~ 0.2 $T_m$, where $T_m$ is melting point). After applying heat transfer, the induced temperature gradient will range from 101 K (= 0.28 $T_m$) to 47 (= 0.13 $T_m$). Such temperature gradient will not make thermal evolution of twinning morphologies, as two atomic configurations before and after applying heat flow shown in Fig. S2. Actually, the temperature range is dependent on the frequency of heat exchange.[1]

3. Phonon spectrum analysis

We further probe the role of VTBs on the reduction of thermal conductivity by analyzing phonon spectrum. Figure S3a shows the phonon frequency distribution of three absolutely different systems, i.e., the bulk one in absence of TBs (Fig. S3b), the one solely containing HTBs (Fig. S3c) and the one primarily containing VTBs (Fig. S3d). The phonon spectrum was obtained by taking the Fourier transformation of auto-correlation function of velocity in *x* component. The vibrational frequency ranges from 0 THz to 40 THz. We find that for bulk system, the most probable frequencies are distributed around 20 THz, as two major peaks shown in Fig. S3b. Thus, the phonons with the mediate-frequency and wavelength dominate the thermal transport. The frequency distribution in system containing HTBs keeps almost the same trend that there are also two peaks, but has little scattering (Fig. S3c). Thus, the existence of HTBs does not affect the heat flux greatly. However, for system containing the high density of VTBs (Fig. S3d), we find that the phonon spectrum shows an obvious change. The phonon spectrum at mediate frequency regime is severely suppressed, which indicates that the contribution of mediate-wavelength phonons on the thermal conductivity will be reduced. Therefore, the present results give the strong evidence that VTBs plays the dominant role on phonons scattering and the heat transport can be blocked by VTBs greatly.

**References for Supplementary Information**

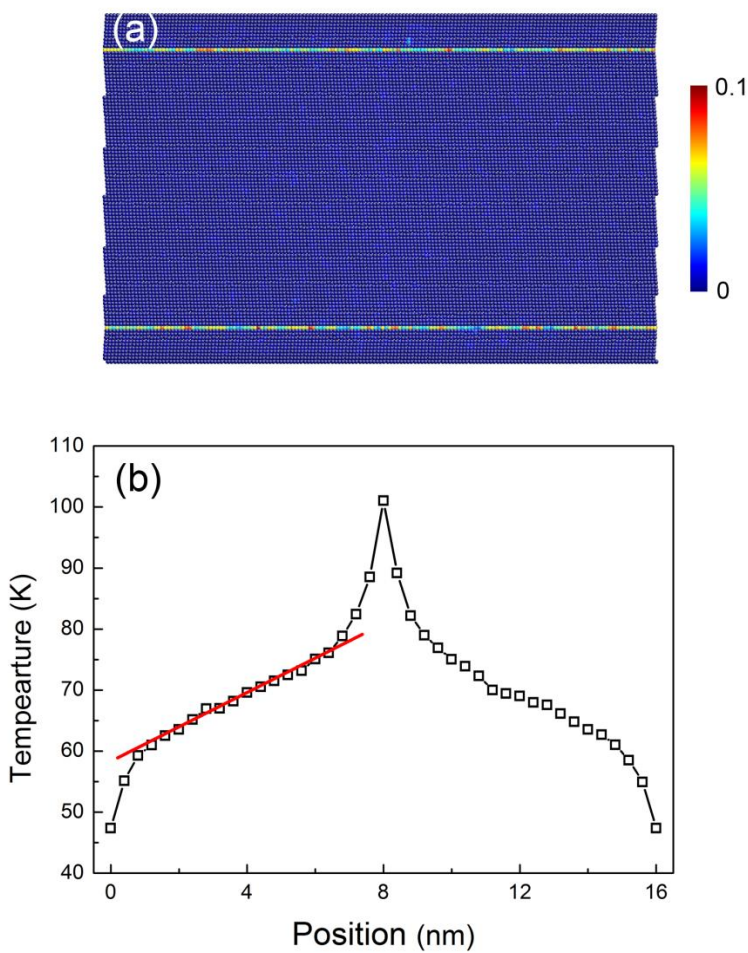

**Figure S1**. Non equilibrium molecular dynamics (NEMD) method. (a) Atomic image of a calculation sample. (b) Temperature profile along *x* direction.

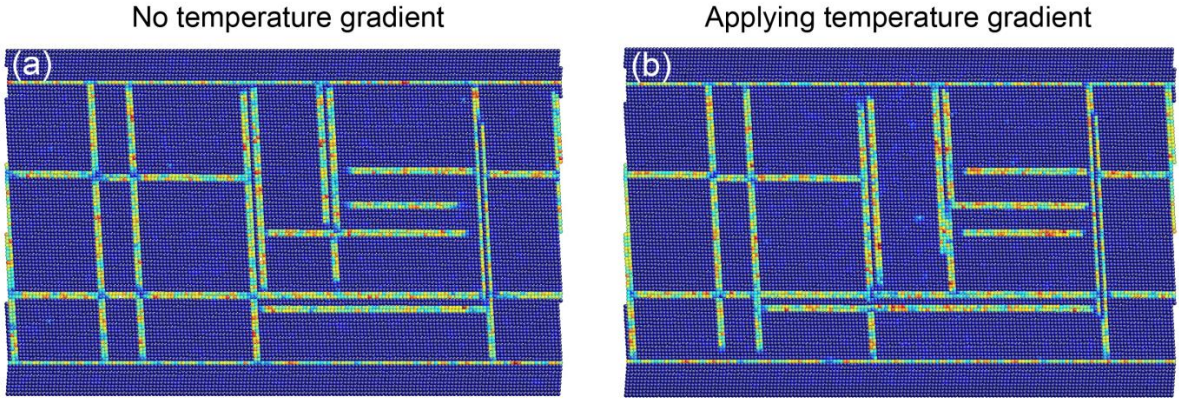

**Figure S2**. The effect of temperature gradient on the morphology of twin patterns. (a) without temperature gradient, (b) with temperature gradient.

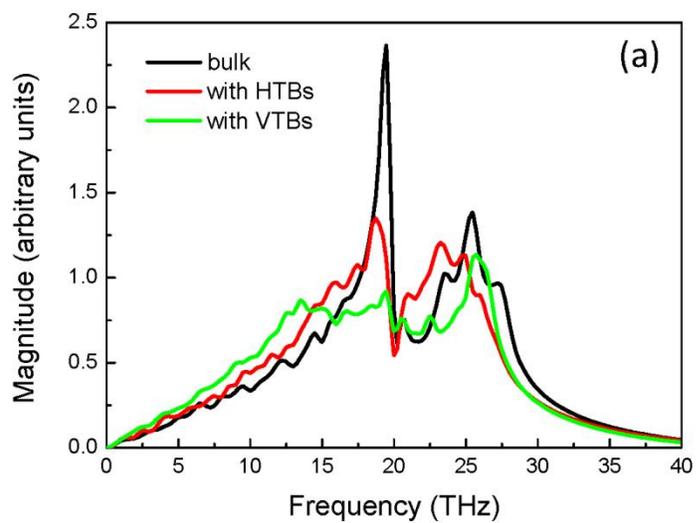
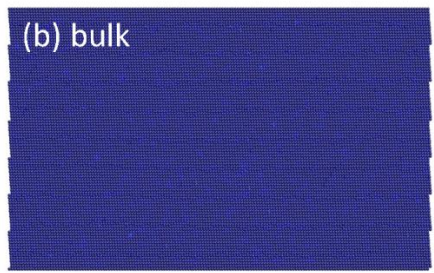
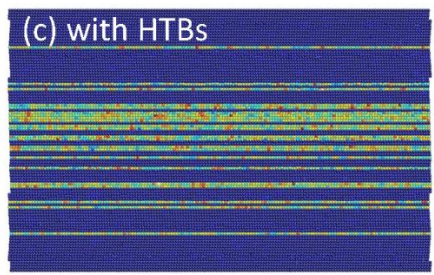
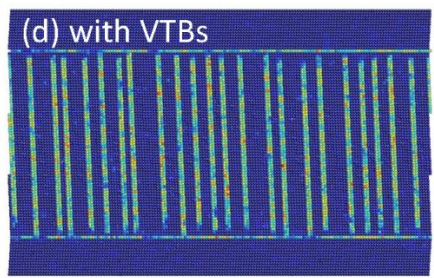

**Figure S3.** (a) Phonon spectrum of three absolutely different systems, as the corresponding atomic configurations shown in (b) bulk, (c) with HTBs and (d) with VTBs.

18